# A modified peer instruction protocol: peer versus teacher's instruction

Boon Leong Lan, Pooi Mee Lim and Patrick W. C. Ho

Electrical and Computer Systems Engineering, School of Engineering, Monash University, 47500 Bandar Sunway, Selangor, Malaysia

Peer Instruction (PI) was introduced by Mazur [1] to help students learn physics concepts during lectures. Besides physics [2,3], PI has also been adopted in other STEM fields [4]. In this approach (Figure 1(a)), students answer a related question individually after a concept has been presented. Before they revote on the same question individually, they are asked to convince others their answer is correct during peer discussion. The percentage of correct answer typically increased after peer discussion [2,3]. However, Smith et al. [5] highlighted that the improvement may be due to copying, not because students actually learned how to reason correctly. To exclude copying, Smith et al. [5] modified Mazur's PI protocol by adding a second question Q2 after the students revote on the first question Q1 (Figure 1(b)). Q2 is 'isomorphic' to Q1, meaning that it requires the application of the same concept but the 'cover story' is different [5]. Here, we simplify Smith et al.'s PI protocol by removing the revote on Q1 (Figure 1(c)). Moreover, our Q1 and Q2 are similar, i.e. the same but some information given is different. Our PI protocol is thus the same as Mazur's, except the pre and post discussion questions are not exactly the same. We replace PI in our protocol by teacher's instruction (TI) to compare the effectiveness of PI with TI for a pair of similar questions involving Lenz's law, using Hake's normalized gain [6] and a statistical test.

The study was conducted in our first-year Physics for Engineering course over five semesters from 2020 to 2021. The pair of similar questions involving the application of Lenz's law is as follows

*The smaller circle in Figure 2 is the cross section of a solenoid. The cross section of the solenoid, and the conducting loop both lie on the page. The magnetic field outside the solenoid is zero. The magnetic field inside the solenoid is uniform, where the magnitude $B_{sol}$ is given by $\mu_0 ni$. The current i in the solenoid is _______ and ________.*

(a) The magnetic flux through the conducting loop is increasing or decreasing?
(b) The current induced in the conducting loop produces a magnetic field. The direction of this magnetic field in the area bounded by the conducting loop is into the page or out of the page?
(c) The induced current in the conducting loop is clockwise (cw) or counterclockwise (ccw)?

The information given for the current in the solenoid is one of four possibilities: cw and decreasing, cw and increasing, ccw and decreasing, or ccw and increasing. For Q1, the current information is not the same for all students (to compel students to learn through peer discussion). Similarly for Q2. For each student, the current information is not the same for Q1 and Q2.

Each question is answered correctly only if all three parts are answered correctly, in this case 1 mark is awarded; otherwise, no mark is given. Before students attempted Q1, they were told that the probability of guessing all three parts correctly is only 0.125. After answering Q1, they were told Q2 will be similar to Q1, and they were asked to learn how to arrive at the answers (which are not revealed at this stage) to Q1 during peer discussion (Semester C, D, E) and teacher's instruction (Semester A, B). As an incentive to do so, Q2 (like Q1) carries a small credit. 5 minutes were allocated for answering each question. The duration for peer discussion was 10 minutes. For teacher's instruction (about 5 minutes), the teacher said the following for each part of Q1 but did not discuss them specifically for any of the four current variations:

(a) The magnetic flux through a loop in a uniform magnetic field is given by $BA$.
(b) Firstly, the direction of the solenoid field is determined by the right curl rule (four fingers curled in the direction of current, thumb points in the field direction). Secondly, according to Lenz's law, if the magnetic flux increases (decreases), the induced field points in the opposite (same) direction to (as) the solenoid field.
(c) The direction of induced current is determined by the right curl rule (thumb points in the field direction, four fingers curled in the direction of current).

Nothing else was mentioned by the teacher beyond the general statements above. These points were already covered separately in the lectures, which is why we administered the pair of questions during the tutorial (problem-solving) session after the lectures. For both PI and TI, the answers to Q1 and Q2 were revealed after Q2 was answered by the students. Both questions were administered as Google forms and the students submitted their answers through the forms.

For TI (Semester A & B, 121 students in total) and PI (Semester C & D & E, 137 students in total), the percentage of correct answer (PCA) for Q1 (see Table 1) is in the lower part of the 35%-70% range proposed by Couch and Mazur [2] for the definition of a challenging question. Thus the pair of similar questions could be classified as highly challenging. Hake's normalized gain, which is a rough measure of the effectiveness of a method of instruction, is defined as [6]

$$\frac{\langle \text{post score} \rangle - \langle \text{pre score} \rangle}{100\% - \langle \text{pre score} \rangle} \quad (1)$$

where <> denotes class average, which in our case is the same as the PCA. The normalized gain (actual gain divided by maximum possible gain) is 0.18 and 0.37 for PI (Semester C & D & E) and TI (Semester A & B), respectively (Table 1). Based on this rough measure, TI is twice as effective as PI for the pair of highly challenging questions. We note that Hake's normalized gain is also consistently lower for PI (see Table 1) compared to TI, based on the data for individual semester.

To assess whether there was a statistically significant difference between the Q1 (pre instruction) and Q2 (post instruction) marks, we used the paired samples Wilcoxon test [7], instead of the paired samples t-test since neither Q1 nor Q2 marks are normally distributed. The test shows that the median of the differences between the paired marks is statistically significantly different from zero (p value = 0.0015) for TI (Semester A & B), but not statistically significantly different from zero (p value = 0.12) for PI (Semester C & D & E). In other words, TI elicited a statistically significant change (since the p value is < 0.05) in the student's paired marks (from Q1 to Q2), but PI did not (p value > 0.05). The conclusion is similar based on the data for individual semester (see the p values in Table 1).

Hake's normalized gain and Wilcoxon's test show that TI is more effective than PI, in our protocol, for the highly challenging pair of similar questions involving Lenz's law. It would be interesting to replicate our pilot study to see if the same conclusion holds. Furthermore, our study could be extended to other pairs of similar questions of different topic and difficulty level (as measured by the PCA for Q1). One could also study whether adding TI after PI in our protocol is more effective than either PI or TI alone between the two similar questions.

**References**

1. E. Mazur, *Peer Instruction: A User's Manual* (Prentice Hall, New Jersey, 1997).

**Figure 1**. Variations of the peer instruction (PI) protocol. (a) Mazur's protocol, (b) Smith et al.'s protocol, and (c) Our protocol. The answer to Q1 is not revealed until after Q2 is answered by the students in protocols (b) and (c), together with the answer to Q2. In this paper, we replace peer discussion in our PI protocol in (c) by teacher's instruction (TI) to compare the effectiveness of PI with TI. Students learn how to arrive at the answer for Q1 from their peers and teacher in PI and TI, respectively.

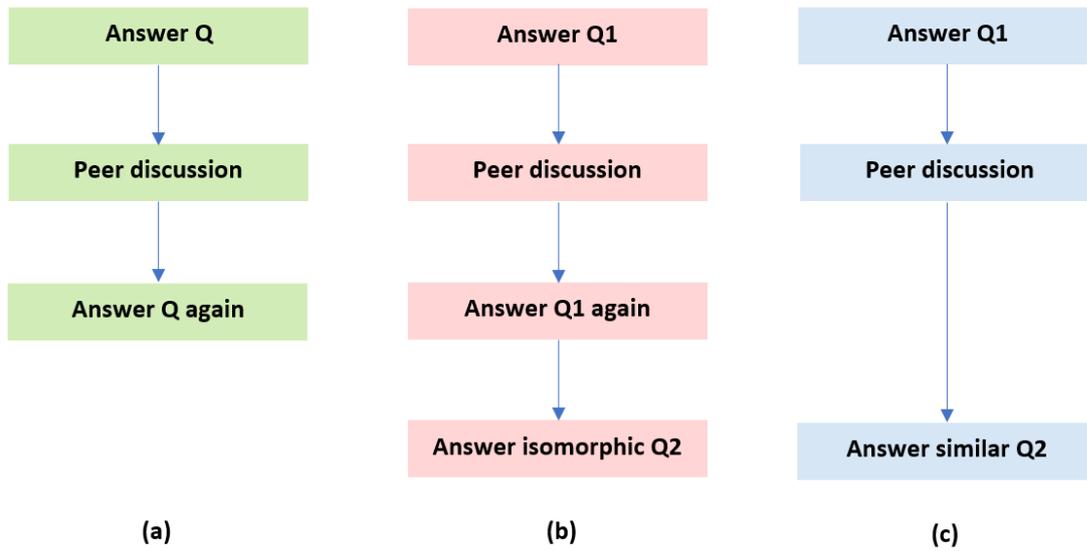

**Figure 2**. Cross section of a solenoid, and a circular conducting loop outside the solenoid for the two similar questions involving Lenz's law.

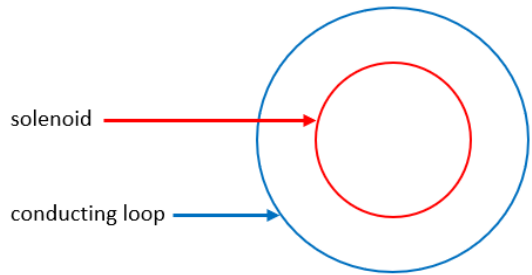

**Table 1**. The percentage of correct answer for each question, Hake's normalized gain, and p value (Wilcoxon test) for peer instruction PI and teacher's instruction TI based on individual semester data and combination of data from different semesters. n is the number of students. The teacher instructor is the same for Semester A and B.

| Instruction | Semester | n | Q1 %correct | Q2 %correct | Hake's norm. gain | p value |
|---|---|---|---|---|---|---|
| TI | A | 95 | 43 | 59 | 0.28 | 0.027 |
| TI | B | 26 | 42 | 81 | 0.67 | 0.0068 |
| TI | A & B | 121 | 43 | 64 | 0.37 | 0.0015 |
| PI | C | 39 | 46 | 59 | 0.24 | 0.306 |
| PI | D | 40 | 40 | 45 | 0.08 | 0.622 |
| PI | E | 58 | 59 | 69 | 0.24 | 0.284 |
| PI | C & D & E | 137 | 50 | 59 | 0.18 | 0.12 |